\documentclass[11pt]{article}

\usepackage{graphicx}
\usepackage{color}
\usepackage{amsmath}
\usepackage{amsfonts}

\newcommand{\ene}{$\displaystyle \frac{\epsilon}{2(2n)}$}
\newcommand{\eno}{$\displaystyle \frac{\epsilon}{2(2n-1)}$}
\newcommand{\evenQ}{$\displaystyle \frac{e^2}{16 \pi^2} \frac{\epsilon^2}{(2n)^2} \int dx$}
\newcommand{\oddQ}{$\displaystyle \frac{e^2}{16 \pi^2} \frac{\epsilon^2}{(2n-1)^2} \int dx$}
\newcommand{\even}{$\displaystyle \frac{g^2}{16 \pi^2} \frac{\epsilon^2}{(2n)^2} \int dx$}
\newcommand{\odd}{$\displaystyle \frac{g^2}{16 \pi^2} \frac{\epsilon^2}{(2n-1)^2} \int dx$}
\newcommand{\evenZ}{$\displaystyle \frac{1}{16 \pi^2} \frac{\epsilon^2}{(2n)^2} \int dx$}
\newcommand{\oddZ}{$\displaystyle \frac{1}{16 \pi^2} \frac{\epsilon^2}{(2n-1)^2} \int dx$}
\newcommand{\li}[1]{$\: \Big( \mbox{Li}_{#1} \left(-e^{-\pi \zeta}\right) - \mbox{Li}_{#1} \left(e^{-\pi \zeta} \right) \Big)$}

\begin{document}

\begin{titlepage}

\begin{flushright}
%SNS-PH/01-14 \\
%UCB-PTH-00/38 \\
%LBNL-47128 \\
\end{flushright}

\vskip 2cm

\begin{center}
{\Large \bf  Muon anomalous magnetic moment in a Calculable Model \\
             with one Extra Dimension}

\vskip 1.0cm

{\bf
Giacomo Cacciapaglia$^{a,}$\footnote{cacciapa@cibs.sns.it},
Marco Cirelli$^{a,}$\footnote{mcirelli@cibs.sns.it},
Giampaolo Cristadoro$^{b,}$\footnote{cristado@cibs.sns.it}
}

\vskip 0.5cm

$^a$ {\it Scuola Normale Superiore and INFN, Piazza dei Cavalieri 7, 
                 I-56126 Pisa, Italy}\\
$^b$ {\it `Enrico Fermi' Dep. of Physics, University of Pisa, via Buonarroti 2, I-56126 Pisa, Italy}

\vskip 1.0cm

\abstract{In the framework of a recently proposed extension of the Standard Model, with $N=1$ SuperSymmetry in 5 dimensions, compactified on $\mathbb{R}^1/\mathbb{Z}_{2} \times \mathbb{Z}'_{2}$, we compute the muon anomalous magnetic moment at one loop to order $(M_W R)^2$, where $R$ is the compactification radius. 
We find the corrections to be small with respect to the SM pure weak contribution and not capable of explaining the present discrepancy between theory and experiment for any sensible value of $R$. 
}

\end{center}

\vspace{8cm}


\end{titlepage}

\setcounter{footnote}{0}

\pagebreak

\section{Introduction}

Models with extra (space) dimensions often fail to produce quantitative predictions for physical observables and, as a consequence, can hardly be ruled out or confirmed by present energy experiments. 
On the other hand, in \cite{BHN} a model was proposed where calculability is achieved for a number of physical quantities. 
In fact, in spite of the non renormalizability of 5 dimensional Yang-Mills theory, they are finite and cut-off independent, owing to the underlying supersymmetric structure.
The insensitivity to ultraviolet physics of such observables is one of the main goals of the model.
The branching ratio of $B \rightarrow X_{s}\gamma$ \cite{BCR} and the Higgs production via gluon fusion \cite{Hgg} have already been computed, while in the present paper we deal with the muon anomalous magnetic moment $a_{\mu}$.

The measurement of $a_{\mu}$ is one of the most stringent tests for ``new physics'' scenarios, thanks to its current impressive precision. 
Moreover, the recent results of E821 experiment at BNL \cite{BNL} have highlighted a discrepancy with the present Standard Model prediction and have therefore made even more interesting the comparison with reliable predictions made by new models.
In particular, attempts have been made in order to study the effect of extra space dimensions \cite{efforts}.

Let us briefly review the present status of the SM prediction \cite{prades}.
The QED contribution has been computed up to order $\alpha^5$ and it gives the core of the experimental value: $a_{\mu}^{QED} = (11\: 658\: 470.0\pm0.3) \cdot 10^{-10}$.
The purely weak terms are also known with a small uncertainty, up to two loops: $a_{\mu}^{weak} = (15.1\pm 0.4)\cdot 10^{-10}$.
On the contrary, hadronic loop corrections provide relevant contributions but also carry the greatest theoretical uncertainty \cite{marciano}.
In fact, since it is quite hard to deal with low energy QCD, experimental inputs are also needed.
Light-by-light scattering yields $a_{\mu}^{lbl} = - (8.6\pm 3.2) \cdot 10^{-10}$ and hadronic vacuum polarization $a_{\mu}^{hvp} = (684.9\pm6.4)\cdot 10^{-10}$.
Summing up all these contributions and uncertainties, the SM prediction is:

$$
a_{\mu}^{SM} = (11\: 659\: 162.0 \pm 7.5)\cdot 10^{-10}
$$
On the other hand, the present world average for the experimental value (including BNL results) is \cite{prades}:

$$
a_{\mu}^{exp} = (11\: 659\: 202.3\pm 15.1) \cdot 10^{-10}
$$
This leads to a discrepancy a bit larger than $2\sigma$:

\begin{equation} \label{dicr}
a_{\mu}^{exp} - a_{\mu}^{SM} = (40.3\pm 16.9)\cdot 10^{-10}
\end{equation}
Actually, this result is not well established.
On the theoretical side, recent results \cite{knecht} \cite{Blokland} claim that the light-by-light contribution has opposite sign, thus reducing the discrepancy to about $1\sigma$.
At the same time, the error on the experimental value should be improved by further results at BNL.

In this work we compute the corrections $\Delta a_{\mu}$ to the anomalous magnetic moment due to the new fields in the theory, up to order $(M_W R)^2$ where $R$ is the compactification radius of the $5^{th}$ dimension.

\section{The model}

In this section we briefly summarize the main features of the model \cite{BHN} we are working with.
It is a 5D theory with N=1 supersymmetry, compactified on $\mathbb{R}^1/\mathbb{Z}_{2} \times \mathbb{Z}'_{2}$, and the gauge group is the Standard Model one: $SU(3)\times SU(2)\times U(1)$. 
The fields content is made up of the vector hypermultiplets and a matter hypermultiplet for each SM matter field $Q$, $U$, $D$, $L$, $E$, $H$, all living in the bulk.

From a 4D point of view, a matter hypermultiplet splits into a pair of chiral supermultiplets with conjugate quantum numbers (property that we indicate with a ${}^c$), while a vector hypermultiplet splits into a vector and a chiral supermultiplet.
Under the orbifolding, global supersymmetry is completely broken {\it \`a la} Scherk-Schwarz, leaving the SM fields as the only zero modes.
Anyhow, restricted local supersymmetric transformations still hold.
The most general Lagrangian, according to this set of symmetries (gauge group, orbifold parities and local supersymmetry), is:

\begin{equation} \label{genlagr}
\mathcal{L} (x, y) = \mathcal{L}_{5} + \delta(y)\: \mathcal{L}_{4} + \delta(y - \pi R/2)\: \mathcal{L}'_{4}
\end{equation}
where $\mathcal{L}_{5}$ is $N=1$ supersymmetric in 5D whereas $ \mathcal{L}_{4}$ and $\mathcal{L}'_{4}$ are 4D lagrangians invariant under different $N=1$ supersymmetries valid on the fixed points of the two parities.
To be consistent with ref. \cite{BHN}, we will write all hypermultiplets in terms of supermultiplets of the 4-dimensional N=1 supersymmetry located on the brane $y=0$.
So, the gauge vector and matter hypermultiplets are:
$$
\big(W_{\rho}, \lambda \big) \oplus \big(\lambda', \Sigma = \frac{1}{\sqrt{2}} (\sigma + i W_5) \big)
$$
$$
\big(\psi_X, \phi_X \big) \oplus \big(\psi_X^c, \phi_X^c \big)
$$
Under the two parities all the fields have definite transformation properties, given in Table \ref{tabella} together with the corresponding eigenfunctions and the spectrum of every tower of Kaluza-Klein states.

\begin{table}[h]
\begin{center}
\begin{tabular}{|c|c|c|c|}
\hline
$A_{\mu}$, $\psi_{M}$, $\phi_{H}$ & $(+, +)$ & $\displaystyle \cos{\frac{2n}{R} y}$ & $\displaystyle \frac{2n}{R}$, $n$ = 0, 1, 2, 3...\\
\hline
$\lambda$, $\phi_{M}$, $\psi_{H}$ & $(+, -)$ & $\displaystyle \cos{\frac{2n-1}{R} y}$ & $\displaystyle \frac{2n-1}{R}$, $n$ = 1, 2, 3...  \\
\hline
$\psi_{\Sigma}$, $\phi_{M}^{c\dagger}$, $\psi_{H}^{c\dagger}$ & $(-, +)$ & $\displaystyle \sin{\frac{2n-1}{R} y}$ & $\displaystyle \frac{2n-1}{R}$, $n$ = 1, 2, 3...\\
\hline
$\phi_{\Sigma}$, $\psi_{M}^{c\dagger}$, $\phi_{H}^{c\dagger}$ & $(-, -)$  & $\displaystyle \sin{\frac{2n}{R} y}$ & $\displaystyle \frac{2n}{R}$, $n$ = 1, 2, 3...\\
\hline
\end{tabular}
\end{center}
\caption{{\small Gauge, Matter and Higgs fields content of the theory with their orbifolding properties}} \label{tabella}
\end{table}

In \cite{SSSZ} it is shown that in the model under consideration the hypercharge current has an anomaly, which is localized on the two branes.
A relevant feature is that, from a 4 dimensional point of view, the integrated anomaly vanishes.
Gauge invariance can however be recovered with a suitable modification of the theory.
The overall consistency of the model modified by the addition of a Chern-Simons term, for instance, is currently under examination.
Such modifications, however, should not affect the calculation presented in this paper, since anomalous contributions only enter at higher order.\\

The terms in the 5D action, that are relevant for the following calculation, are:

\begin{equation} \label{lagr}
\begin{array}{rcl}
S & = \int d^5 x &\left\{ -\displaystyle \frac{1}{4} \mathbf{Tr} F_{MN} F^{MN} - \mathbf{Tr} \left[ i \lambda \sigma^{\rho} \mathcal{D}_{\rho} \overline{\lambda} - i \lambda' \sigma^{\rho} \mathcal{D}_{\rho} \overline{\lambda}' \right] + i \psi_{X} \sigma^{\rho} \mathcal{D}_{\rho} \overline{\psi}_{X} + i \psi_{X}^{c} \sigma^{\rho} \mathcal{D}_{\rho} \overline{\psi}^{c}_{X} + \right.\\
 && + (\mathcal{D}_{M}\phi_{X})^{\dagger}(\mathcal{D}^{M}\phi_{X}) + (\mathcal{D}_{M}\phi_{X}^{c})^{\dagger}(\mathcal{D}^{M}\phi_{X}^{c}) + \displaystyle \sqrt{2} g_{(5)} \left[ \phi_{X}^{\dagger} \lambda \psi_{X} - \psi^{c}_{X} \lambda \phi_{X}^{c \dagger} + h.c. \right] + \\
&& \displaystyle +\sqrt{2} g_{(5)}  \left[\psi_{X}^{c} \lambda' \phi_{X} + \phi_{X}^{c} \lambda' \psi_{X} + \psi_{X}^{c} \Sigma \psi_{X} + h.c. \right] +\\
&& + \displaystyle \delta (y - \frac{\pi}{2} R) \lambda_{\mu} \left[\psi_{L} \psi_{E} \phi_{H^{0}}^{\dagger} + \psi_{L} \psi^{c}_{H^{0}} \phi^{c\dagger}_{E} + \psi_{E} \psi^{c}_{H^{0}} \phi^{c\dagger}_{L}+ \right. \\
&& \displaystyle \left. \left. - ( \psi_{\nu} \psi_{E} \phi_{H^{+}}^{\dagger} + \psi_{\nu} \psi^{c}_{H^{+}} \phi^{c\dagger}_{E} + \psi_{E} \psi^{c}_{H^{+}} \phi^{c\dagger}_{\nu} ) \right] \right\}
\end{array}
\end{equation}
where the fermionic fields are Weyl spinors, the gauge fields $W_{M}$, $\lambda$, $\lambda'$ and $\sigma$ contains the group generators, e.g. $\lambda \equiv \lambda^{a} T^{a}$ (the generators being normalized according to $\mathbf{Tr}(T^{a}T^{b})=\delta^{ab}$) and the index $X$ can run over $E$, $L$, $H$. 
The gauge coupling $g$ is intended to be the $U_{Y}(1)$ coupling $g'$ or the $SU_{L}(2)$ coupling $g$ according to the gauge fields involved.

From now on, for the four gauge hypermultiplets we use the notation:

\begin{eqnarray*}
QED & \big( A_{\rho}, \tilde{\gamma}, \tilde{\gamma}_{c}, \Sigma_{\gamma} \big) &\\
Z's & \big( Z_{\rho}, \tilde{z}, \tilde{z}_{c}, \Sigma_{z} \big) &\\
W's & \big( W_{\rho}^{\pm}, \tilde{w}^{\pm}, \tilde{w}_{c}^{\pm}, \Sigma_{w}^{\pm} \big)&
\end{eqnarray*}
(we remind that, also in this case, the subscribed $c$ only means that the second spinors have conjugate quantum numbers).

\subsection{Mass Eigenstates}

The brane interactions introduce mass mixing among the states of muon and smuon KK towers: since it will be useful for our calculation to work with the mass eigenstates, we proceed to determine eigenvalues and eigenvectors from the very beginning (see e.g. the discussion in \cite{BCR}).
For the fermions, the mass eigenvalues turn out to be:

$$
m_{0} = m \equiv  \frac{\epsilon}{R}, \qquad  m_{n}^{\pm} = m \pm \frac{2n}{R} \qquad  (n=1,2,...)
$$
corresponding to mass eigenvectors whose Weyl spinor components we denote by $(\mu_{0},\mu_{0}^{c})$ and $(\mu_{n}^{\pm},\mu_{n}^{c\pm})$. Thus $m$ is the mass of the ordinary muon.
For the scalars:

$$
 \qquad \tilde{m}_{n}^{\pm} = \frac{(2n-1)}{R} \pm m \qquad (n=1,2,...)
$$
corresponding to eigenvectors $\Phi_{n}^{\pm}$ and $\Phi_{n}^{c\pm}$.

The interaction eigenstates\footnote{Note that in this paper the brane lagrangians $\mathcal{L}_4$ and $\mathcal{L}'_4$ are reversed respect to ref \cite{BHN}. For more details, see appendix A in \cite{Hgg}.} are expressed in terms of the mass eigenstates (at order $\epsilon$) by
\begin{eqnarray}
\psi_{L,0} & = & \displaystyle \mu_{0} + \epsilon \sum_{n} \frac{1}{2n} \big(\mu_{n}^{+} - \mu_{n}^{-} \big) \label{psiLzero}\\
\psi_{L,l} & = & \displaystyle \frac{1}{\sqrt{2}} \left\{ \left( 1 + \mbox{\ene} \right) \mu_{l}^{+} + \left( 1 - \mbox{\ene} \right) \mu_{l}^{-} + \epsilon \sum_{n} \kappa_{nl} \big( \mu_{n}^{+} - \mu_{n}^{-} \big) \right\}  \label{psiLenne}\\
\psi^{c}_{E,l} & = & \displaystyle \frac{1}{\sqrt{2}} \left\{ \left( 1 - \mbox{\ene} \right) \mu_{l}^{+} - \left( 1 + \mbox{\ene} \right) \mu_{l}^{-} - \epsilon \sum_{n} \kappa_{ln} \big( \mu_{n}^{+} + \mu_{n}^{-} \big) \right\} - \frac{\sqrt{2}}{2l} \epsilon\: \mu_{0} \label{psiEc}\\
  \nonumber
\end{eqnarray}

\begin{eqnarray}
\phi_{L,l} & = & \displaystyle \frac{1}{\sqrt{2}} \left\{ \left( 1 + \mbox{\eno} \right) \Phi_{l}^{+} + \left( 1 - \mbox{\eno} \right) \Phi_{l}^{-} + \epsilon \sum_{n} J_{ln} \big(\Phi_{n}^{+} - \Phi_{n}^{-} \big) \right\} \label{phiEc} \\
\phi^c_{E,l} & = & \displaystyle \frac{1}{\sqrt{2}} \left\{ - \left( 1 - \mbox{\eno} \right) \Phi_{l}^{+} + \left( 1 + \mbox{\eno} \right) \Phi_{l}^{-} + \epsilon \sum_{n} J_{nl} \big(\Phi_{n}^{+} + \Phi_{n}^{-} \big) \right\} \label{phiL}
\end{eqnarray}
with $\kappa_{ln} = \frac{2 (2l)}{(2l)^2 - (2n)^2}$ and $J_{ln} = \frac{2 (2l-1)}{(2l-1)^2 - (2n-1)^2}$, and zero for $n=l$.
Expressions which are identical to (\ref{psiLzero}), (\ref{psiLenne}), (\ref{psiEc}) hold for ($\psi_{E,0}$, $\psi_{E,n}$, $\psi^{c}_{L,n}$) if ($\mu_{0}$, $\mu_{n}^{\pm}$) are replaced by ($\mu_{0}^{c}$, $\mu_{n}^{c \pm}$); for ($\phi_{L}^{c}$, $\phi_{E}$) it is sufficient to replace $\Phi_{n}^{\pm}$ with $\Phi_{n}^{c \pm}$ in equations (\ref{phiEc}), (\ref{phiL}).

\subsection{Couplings to the mass eigenstates}

Starting from eq. (\ref{lagr}), one has to integrate out the 5$^{th}$ dimension $y$ and rewrite the interaction vertices in terms of the mass eigenstates (\ref{psiLzero})-(\ref{phiL}).
Taking into account momentum conservation along the 5$^{th}$ direction, $p_5$, the bulk interactions give rise to several trilinear vertices.
The muon masses and Higgs vertices, on the other hand, being located on the brane, violate $p_5$-conservation and are proportional to $\epsilon$.
As we are interested in $\mathcal{O} (R^2)$ contributions, only graphs where the internal lines carry the same index $n$, in terms of the mass eigenstates, have to be taken into account.
This simplifies the calculation of the relevant vertices.
As an explicit example, the interaction terms between KK photons and the various components in the muon towers, in the mass eigenstate basis, are:

\begin{eqnarray*}
e \overline{\psi}_{L,0} \overline{\sigma}_{\rho} \psi_{L,n} A^{\rho}_{n} &=& e \overline{\mu}_{0} \overline{\sigma}_{\rho} \frac{1}{\sqrt{2}} \left\{ \left( 1 - \mbox{\ene}\right) \mu^{+}_{n} + \left( 1 + \mbox{\ene}\right) \mu^{-}_{n} \right\} A_{n}^{\rho}\\
\frac{e}{\sqrt{2}} \overline{\psi}^{c}_{E,2n} \overline{\sigma}_{\rho} \psi_{E,n}^{c} A_{n}^{\rho} &=& - e \mbox{\ene} \overline{\mu}_{0} \overline{\sigma}_{\rho} \frac{1}{\sqrt{2}} \big\{ \mu^{+}_{n} -  \mu^{-}_{n} \big\} A_{n}^{\rho}
\end{eqnarray*}
In this particular case, summing up the two contributions, the couplings of order $\epsilon$ vanish, so that the net result is:

\begin{equation} \label{vertexKKphotons}
e \overline{\mu}_{0} \overline{\sigma}_{\rho} \frac{1}{\sqrt{2}} \left(\mu^{+}_{n} + \mu^{-}_{n} \right) A^{\rho}_{n}
\end{equation}
This allows to recognize $e/\sqrt{2}$ as the correct coupling with a KK photon.

On the other hand, the interactions of the zero mode gauge bosons are still diagonal in $n$ and have a universal coupling, e.g. $e = e_{(5)}/\sqrt{2 \pi R}$ for the photon.
Note that this relation holds for all gauge couplings, so in the following $g = g_{(5)}/\sqrt{2 \pi R}$ is the 4D gauge coupling, while $s=\sin{\theta_W}$ and $c = \cos{\theta_W}$ are functions of the usual SM Weinberg angle.

\section{Calculation of $a_{\mu}$}

The effective Lagrangian term which fixes the notation for the anomalous magnetic moment $a_{\mu}=(g_{\mu}-2)/2$ is:

\begin{equation}
\mathcal{L} = \frac{i e}{2 m}\: a_{\mu}\: \big(\overline{\mu} \sigma_{\rho \sigma} F^{\rho \sigma} \mu \big)
\end{equation}

\begin{figure}[th]
\begin{center}
\includegraphics[width=13cm]{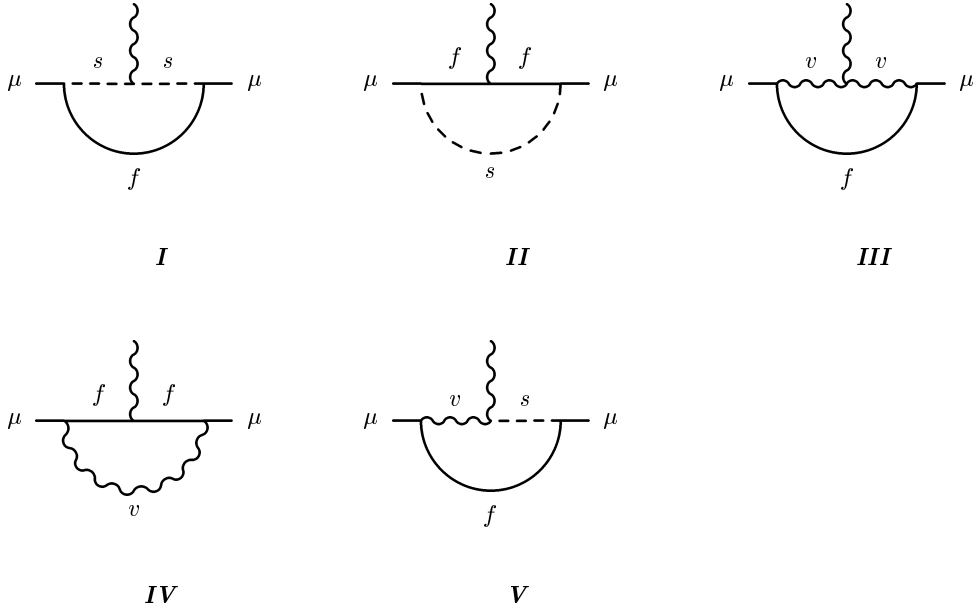}
\caption{{\small General patterns of graphs contributing to the $(g-2)$.}} \label{graphsmu}
\end{center}
\end{figure}

In the Feynman--'t Hooft gauge ($\xi =1$, see below), one can identify the five general types of diagrams in fig. \ref{graphsmu}, each giving a contribution $\Delta a_{\mu}$ whose explicit expressions are listed in app. A.

As in the following we will distinguish between diagrams with chirality flip on the external or internal fermionic line, in eq. (\ref{graphI})-(\ref{graphIV}) it is easy to trace back the external cross contribution and the internal cross one (the latter proportional to $m_f/m$).
On the other hand, in graph \emph{V} the chirality flip is provided by the only scalar interaction with the fermion.

\subsection{QED}

In this section we consider all the diagrams involving QED hypermultiplet fields. 
The one loop Standard Model contribution \cite{schwinger} simply comes from graph \emph{IV} with a massless photon in the loop, with $g_{+} = g_{-} = e$:

\begin{equation}
a_{\mu}^{\gamma} = \frac{\alpha}{2 \pi}
\end{equation}

\subsubsection{KK vectors}

The interaction lagrangian for KK photons has been displayed in (\ref{vertexKKphotons}) and the graphs are of type \emph{IV}. 
As a general rule, dealing with the diagrams involving KK fields, it is useful to separate the external and internal cross contributions. 
This is motivated by the fact that in the latter case it is necessary to use couplings and loop masses at order $\epsilon$ while in the former $\mathcal{O}(1)$ is enough.
The following sketched diagrams are drawn in terms of Weyl fermions and the emitted zero-mode photon is implicitly attached to any internal charged line. 
The states in the loop are KK modes of the field indicated, always carrying the same index $n$ for the reasons explained above.
The  resulting $\Delta a_{\mu}$, for a given internal KK state, are:

\begin{eqnarray}
\parbox[c]{5cm}{\includegraphics[width=4cm]{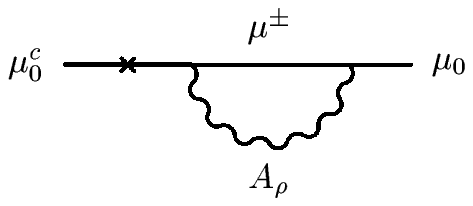}} &\Rightarrow & \mbox{\evenQ} \: \big[4 x (1-x)^2 - 8 x (1-x)\big]\\
\parbox[c]{5cm}{\includegraphics[width=4cm]{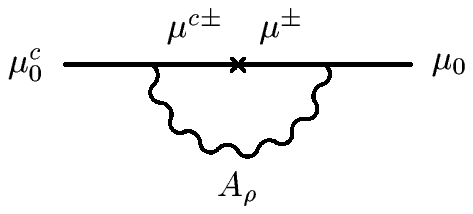}} & \Rightarrow & \mbox{\evenQ}  \:\big[ 8 x (1-x) - 16 x (1-x)^2 \big]
\end{eqnarray}
and the trivial integration over the Feynman parameter $x$ is left explicit in order to discuss supersymmetric cancellations (see below).

\subsubsection{Scalars}

The gauge scalar field $A_{5}$ mixes mode by mode with the massive vector $A_{\rho}$. 
Including also the mass term for the gauge scalar $\sigma_{\gamma}$, the mass lagrangian, for each mode $n$, is:

\begin{equation}
\label{LmassQED}
\mathcal{L}_{mass} = - \partial_{\rho} A^{\rho} \left( \frac{2n}{R} A_{5} \right) - \frac{1}{2} \left( \frac{2n}{R} \right)^2 \sigma_{\gamma}^2 - \frac{1}{2 \xi} \left( \partial_{\rho} A^{\rho} - \xi \frac{2n}{R} A_{5} \right)^2
\end{equation}
where the $R_{\xi}$ gauge fixing term is there to cancel the mixing terms.
It is convenient to work in the $\xi = 1$ gauge: we obtain a photon propagator with no longitudinal modes and two scalars $\sigma_{\gamma}$ and $A_{5}$  (with equal mass) that can be combined in the complex scalar $\Sigma_{\gamma}$. 
The interaction terms are then:

\begin{equation}
\sqrt{2} e\: \big( \psi_{E}^{c} \Sigma_{\gamma} \psi_{E} + \psi_{L}^{c} \Sigma_{\gamma} \psi_{L} + h.c.\big)
\end{equation}
so that at the relevant order in $\epsilon$, only an external cross graph is possible. 
Equivalently, in terms of the real scalar fields it happens that the internal cross diagram with an $A_{5}$ field is cancelled out by the same diagram with a $\sigma_{\gamma}$.
The external cross diagram is of type \emph{II} so that the resulting $\Delta a_{\mu}$ is

\begin{equation}
\parbox[c]{5cm}{\includegraphics[width=4cm]{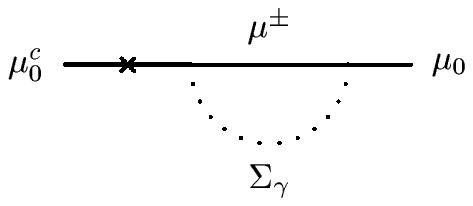}} \Rightarrow \mbox{\evenQ}  \: \big[4 x (1-x)^2 \big] 
\end{equation}

\subsubsection{Fermions}

The gauge fermions $\tilde{\gamma}$ and $\tilde{\gamma}_{c}$ are mixed by the kinetic term along the 5$^{th}$ direction. 
They are involved in the following interaction terms

\begin{eqnarray}
&&  \displaystyle - \sqrt{2} e \mu_{0} \tilde{\gamma}_{n} \frac{1}{\sqrt{2}} \left\{ \left( \Phi^{\dagger -}_{n} - \Phi^{\dagger +}_{n} \right) + \frac{\epsilon}{2n-1} \left( \Phi^{\dagger -}_{n} + \Phi^{\dagger +}_{n} \right) \right\} \nonumber\\
&& +  \sqrt{2} e \mu_{0}^{c} \tilde{\gamma}_{c, n} \frac{1}{\sqrt{2}} \left\{ \left( \Phi_{n}^{+} + \Phi_{n}^{-} \right) + \frac{\epsilon}{2n-1} \left( \Phi_{n}^{+} - \Phi_{n}^{-} \right) \right\}
\end{eqnarray}
and in similar expressions (replacing $\Phi$ with $\Phi^{c}$) for the interactions of $\mu^{c}_{0}$ with $\tilde{\gamma}_{n}$ and $\mu_{0}$ with $\tilde{\gamma}_{c,n}$.\\
From graph \emph{II}, one gets

\begin{equation}
\left. \parbox[c]{5cm}{\includegraphics[width=4cm]{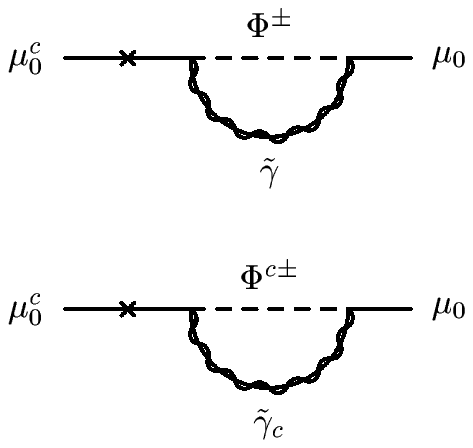}}\right\}  \Rightarrow  - \mbox{\oddQ} \: \big[4 x (1-x)^2\big]  \times 2 
\end{equation}
\begin{equation}
 \left. \parbox[c]{5cm}{\includegraphics[width=4cm]{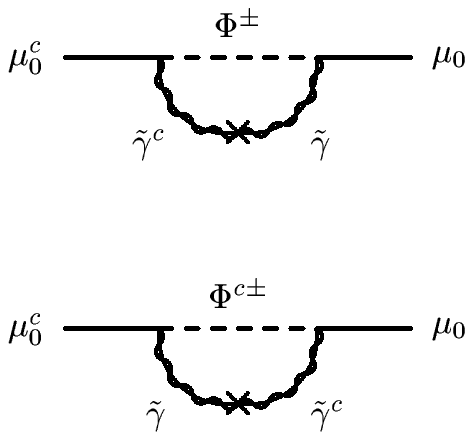}}\right\}  \Rightarrow   \mbox{\oddQ} \: \big[8 x (1-x)^2 \big]  \times 2
\end{equation}

\subsubsection{Total QED contribution}

First of all, in order to test the computation, we consider the supersymmetric limit in which both fermions and bosons have the same parities and therefore the same bulk mass.
In this case, we find that the total contributions of the KK modes vanishes, as expected.
Of course, the appearance of zero modes for the matter and gauge partners also cancels the SM result.

Then, we collect all the QED contributions and we obtain:

\begin{equation} \label{finalqed}
\Delta a^{QED}_{\mu} = \frac{e^2}{16 \pi^2}  (m R)^2 \frac{\pi^2}{18} 
\end{equation}

\subsection{Neutral Weak sector}

In this section we consider all the diagrams involving $Z$ hypermultiplet fields and neutral Higgs. We define

\begin{displaymath}
\begin{array}{lcl}
 g_{L}&=&\displaystyle \frac{1}{c} \left( -\frac{1}{2} + s^2 \right)\: g\\
 g_{R}&=& - \displaystyle \frac{s^2}{c} \: g
\end{array}
\end{displaymath}
The one loop SM result \cite{ACM} comes from graph \emph{IV} with a zero mode $Z$ boson and couplings $g_{-}=g_{L}$, $g_{+}=-g_{R}$:

\begin{equation}
a_{\mu}^{Z} = \frac{g^2}{48 \pi^2} \frac{m^2}{M_{Z}^2} \frac{4s^4-2s^2-1}{c^2}
\end{equation}

The discussion for the KK modes mimics the QED case, thus in the following we only point out the differences.

\subsubsection{KK vectors}
The interaction lagrangian consists in

\begin{equation}
 \overline{\mu}_{0} \overline{\sigma}_{\rho} \frac{1}{\sqrt{2}} \big\{ j^{+}_{L} \mu_{n}^{+} + j^{-}_{L} \mu_{n}^{-} \big\} Z^{\rho}_{n} +  \overline{\mu}^{c}_{0} \overline{\sigma}_{\rho} \frac{1}{\sqrt{2}} \big\{ j^{+}_{R} \mu_{n}^{c+} + j^{-}_{R} \mu_{n}^{c-} \big\} Z^{\rho}_{n} + h.c.
\end{equation}
where:

\begin{equation}
\begin{array}{c}
j_{L}^{+}  =  g_{L} + \mbox{\ene}\: \big(g_{L} + g_{R}\big) \qquad j_{R}^{+}  =  - g_{R} - \mbox{\ene}\:  \big(g_{R} + g_{L}\big) \\
j_{L}^{-}  =  g_{L} - \mbox{\ene}\: \big(g_{L} + g_{R}\big) \qquad  j_{R}^{-}  =  - g_{R} + \mbox{\ene}\: \big(g_{R} + g_{L}\big)       
\end{array}
\end{equation}

The two $\Delta a_{\mu}$ then amount to

\begin{equation} \label{eczeta}
\parbox[c]{5cm}{\includegraphics[width=4cm]{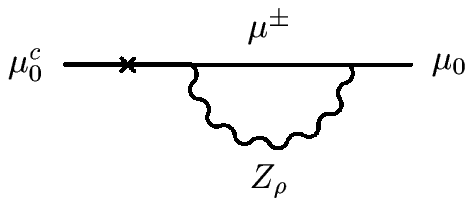}} \Rightarrow \mbox{\evenZ} (g_{L}^2 + g_{R}^2)\: \big[2 x (1-x)^2 - 4 x (1-x)\big] 
\end{equation}

\begin{multline} \label{iczeta}
\parbox[c]{5cm}{\includegraphics[width=4cm]{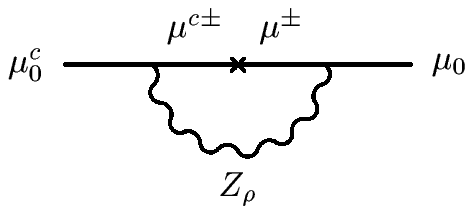}} \Rightarrow - \mbox{\evenZ} \left[ \left( g_{L} g_{R} + \frac{1}{2} (g_{L} + g_{R})^2 \right) 8 x (1-x) \right. \\
 - g_{L} g_{R} 16 x (1-x)^2 \bigg]
\end{multline}

\subsubsection{Scalars}

With respect to the QED case (\ref{LmassQED}), the lagrangian for scalar fields is a bit more complicated, involving also mixing terms with $\varphi_{H^{0}}$, the imaginary part of $\phi_{H^{0}}$. 
With the gauge fixing piece already included, the lagrangian for each mode $n$ is:

\begin{multline} \label{Zscalarsmass}
\mathcal{L}_{mass} =  -  \partial_{\rho} Z^{\rho} \left( \frac{2n}{R} Z_{5} - M_{Z} \varphi_{H^{0}} \right) - \frac{1}{2} \left[ \frac{2n}{R} \varphi_{H^{0}} - M_{Z} Z_{5} \right]^2 - \frac{1}{4} M_{Z}^2 \varphi_{H^{0}}^2 - \frac{1}{2} \left( \frac{2n}{R} \right)^2 \sigma_{Z}^{2}\\ - \frac{1}{2 \xi} \left[ \partial_{\rho} Z^{\rho} - \xi \left(\frac{2n}{R} Z_{5} - M_{Z} \varphi_{H^{0}} \right)\right]^2 
\end{multline}
Vector-scalar mixings are again cancelled and, for $\xi = 1$, there are no off-diagonal terms $\varphi_{H^{0}} Z_{5}$, so the three real scalars $\sigma_{Z}$, $Z_{5}$, $\varphi_{H^{0}}$ are the mass eigenstates. 
The first two combine in the complex field $\Sigma_{Z}$ in terms of which, at the relevant order in $\epsilon$, only the external cross graph is possible:

\begin{equation} \label{ecszeta}
\parbox[c]{5cm}{\includegraphics[width=4cm]{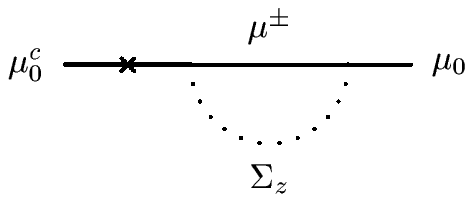}} \Rightarrow \mbox{\evenZ} (g_{L}^2 + g_{R}^2)\: 2x (1-x)^2
\end{equation}

\subsubsection{Fermions}

The mixing matrix for the four neutralinos $\tilde{z}$, $\tilde{z}_{c}$, $\psi_{H^{0}}$, $\psi_{H^{0}}^{c}$, for each mode $n$, is:

\begin{equation} \label{neutralinosmass}
\mathcal{L}_{mass} = \left( \psi_{H^{0}} , \tilde{z} \right) \left( \begin{array}{cc}
\frac{2n-1}{R} & M_{Z} \\
- M_{Z} & \frac{2n-1}{R}
\end{array} \right) \left( \begin{array}{c}
\psi_{H^{0}}^{c} \\
\tilde{z}_{c}
\end{array} \right)
\end{equation}

They are involved in the following interaction terms:

\begin{eqnarray}
&&\displaystyle  \sqrt{2}  \mu_{0} \tilde{z}_{n} \frac{1}{\sqrt{2}} \left\{ g_{L} (\Phi_{n}^{\dagger -} - \Phi_{n}^{\dagger +} ) + (g_{L} - g_{R} ) \mbox{\eno} (\Phi_{n}^{\dagger -} + \Phi_{n}^{\dagger +} ) \right\}+ \nonumber \\ 
&& \displaystyle + \sqrt{2}  \mu^{c}_{0} \tilde{z}_{n} \frac{1}{\sqrt{2}} \bigg\{ g_{L} \leftrightarrow g_{R} \quad \textrm{and} \quad \Phi \leftrightarrow \Phi^{c} \bigg\} + \nonumber\\
&&\displaystyle + \sqrt{2}  \mu^{c}_{0} \tilde{z}^{c}_{n} \frac{1}{\sqrt{2}} \left\{ g_{R} (\Phi_{n}^{-} + \Phi_{n}^{+} ) - (g_{R} - g_{L} ) \mbox{\eno} (\Phi_{n}^{-} - \Phi_{n}^{+} ) \right\} +\\
&&\displaystyle+ \sqrt{2}  \mu_{0} \tilde{z}^{c}_{n} \frac{1}{\sqrt{2}} \bigg\{ g_{L} \leftrightarrow g_{R} \quad \textrm{and} \quad  \Phi \leftrightarrow \Phi^{c} \bigg\} + \nonumber\\
&&\displaystyle + 2 \frac{\epsilon}{v} \left\{ \mu_{0} \frac{1}{\sqrt{2}} (\Phi_{n}^{-} + \Phi_{n}^{+} ) + \mu^{c}_{0} \frac{1}{\sqrt{2}} (\Phi_{n}^{c-} + \Phi_{n}^{c+} ) \right\} \psi_{H,n}^{c} \nonumber
\end{eqnarray}

The external cross graphs are readily computed:

\begin{equation} \label{eczino}
\left. \parbox[c]{5cm}{\includegraphics[width=4cm]{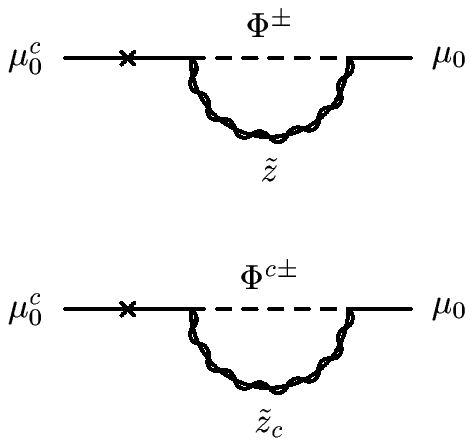}} \right\} \Rightarrow - \mbox{\oddZ} (g_{L}^2 + g_{R}^2 )\: 2 x (1-x)^2 \times 2
\end{equation}
and it is possible to write three internal cross diagrams, using also the off-diagonal propagator in eq. (\ref{neutralinosmass}):

\begin{eqnarray}
\parbox[c]{4.5cm}{\includegraphics[width=4cm]{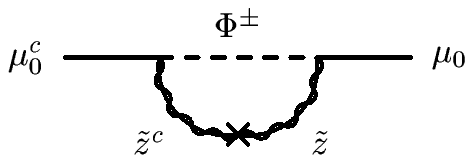}} &\Rightarrow& - \mbox{\oddZ} \left[ (g_{L}^2 - g_{R}^2) 2x (1-x) +  g_{L} g_{R} 8 x (1-x)^2 \right] \label{iczino1}\\
\parbox[c]{4.5cm}{\includegraphics[width=4cm]{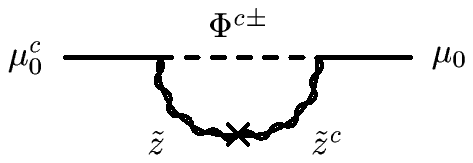}} &\Rightarrow& - \mbox{\oddZ} \left[ (g_{R}^2 - g_{L}^2) 2x (1-x) +  g_{L} g_{R} 8 x (1-x)^2 \right] \label{iczino2}\\
\parbox[c]{4.5cm}{\includegraphics[width=4cm]{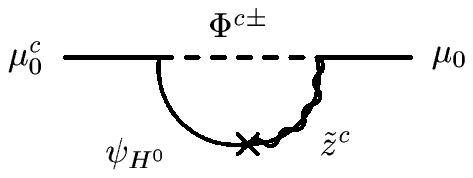}} & \Rightarrow & - \mbox{\oddZ} \frac{g}{c} (g_L + g_R)\: 4 x (1-x) \label{ichiggsino0}
\end{eqnarray}

\subsubsection{Total neutral weak contribution}

Again, as in case of QED, one can check that in the supersymmetric limit, discussed above, the contribution of the KK modes vanishes. 
Moreover, in the limit $g_{L} = -g_{R} = e$, the QED results are recovered.

Then, we sum up all the results of the neutral weak sector and obtain:  

\begin{equation} \label{finalZ}
\Delta a^{Z}_{\mu} = \frac{g^2}{16 \pi^2} (m R)^2 \frac{1}{c^2} \left( \frac{1}{4} - \frac{s^2}{3} + \frac{2 s^4}{3} \right) \frac{\pi^2}{12}
\end{equation}

\subsection{Charged Weak sector}

In this section we consider all the diagrams involving $W^{\pm}$ hypermultiplets fields and charged Higgs. 
The SM result \cite{brodsky} comes from graph \emph{III}, where the loop contains a zero mode $W$ boson and a neutrino and $g_{+} = g/\sqrt{2}$, $g_{-} = 0$:

\begin{equation}
a_{\mu}^{W} = \frac{5g^2}{96 \pi^2} \frac{m^2}{M_{W}^2}
\end{equation}

\subsubsection{KK vectors}

The interaction terms involving the KK vectors are quite simple
\begin{equation}
 \frac{g}{\sqrt{2}} W^{\rho}_{n}\: \mu_{0} \sigma_{\rho} \overline{\psi}_{\nu,n} - \frac{g}{2\sqrt{2}} \frac{\epsilon}{2n} W^{\rho}_{n}\: \mu_{0}^{c} \sigma_{\rho} \overline{\psi}_{\nu,n}
\end{equation}
and the external and internal cross diagrams are readily computed using graph \emph{III}:

\begin{eqnarray} 
\parbox[c]{5cm}{\includegraphics[width=4cm]{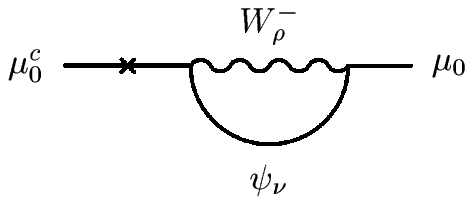}} &\Rightarrow& - \mbox{\even}\:  (1-x)^2 (x -3/2) \label{ecw}\\
\parbox[c]{5cm}{\includegraphics[width=4cm]{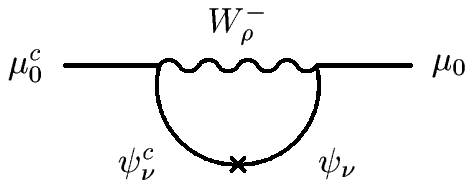}} &\Rightarrow& \mbox{\even}\: \frac{3}{2} (1-x)^2 \label{icw}
\end{eqnarray}

There also are diagrams of type \emph{V}. The only nonvanishing one has a Higgs scalar, since the contributions of $W_{5}$ and $\sigma_{W}$ cancel out:

\begin{equation} \label{higgs}
\parbox[c]{5cm}{\includegraphics[width=4cm]{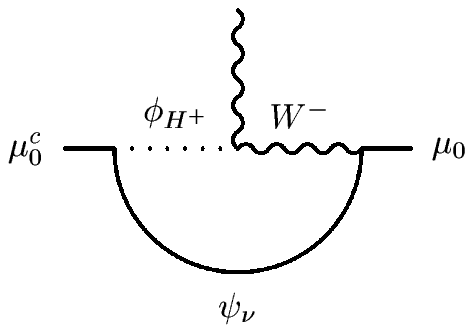}} \Rightarrow \mbox{\even}\: (1-x)^2
\end{equation}

\subsubsection{Scalars}

The gauge fixing procedure works as in the $Z$'s case: the interaction lagrangian terms only allow for an external cross contribution:

$$
g \big(\psi_{\nu}^{c} \Sigma^{+} \psi_{L} + \psi_{L}^{c} \Sigma^{-} \psi_{\nu} + h.c.\big)
$$

\begin{equation} \label{ecsw}
\parbox[c]{5cm}{\includegraphics[width=4cm]{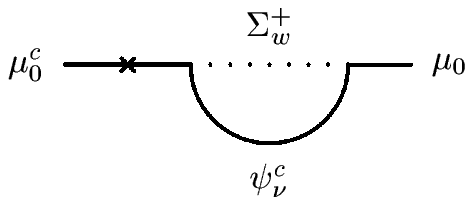}} \Rightarrow - \mbox{\even}\: x (1-x)^2
\end{equation}

\subsubsection{Fermions}

The charginos $\tilde{w}^{-}$, $\tilde{w}^{-}_{c}$, $\psi_{H^{+}}$, $\psi_{H^{+}}^{c}$  mix for each $n$ according to:

\begin{equation} \label{charginosmass}
\mathcal{L}_{mass} =  \left( \psi_{H^{+}}, \tilde{w}^{-}_{c} \right) \left( \begin{array}{cc}
\frac{2n-1}{R} & \sqrt{2} M_{W}\\
-\sqrt{2} M_{W} & \frac{2n-1}{R}
\end{array} \right) \left( \begin{array}{c}
\psi_{H^{+}}^{c}\\
\tilde{w}^{-}
\end{array} \right) - \frac{2n-1}{R} \tilde{w}_{c}^{+} \tilde{w}^{+}
\end{equation}
They enter the following interaction terms

\begin{eqnarray} 
&& \displaystyle  g \phi_{\nu, n}^{\dagger} \tilde{w}_{n}^{+} \mu_{0} - g \mbox{\eno} \phi_{\nu,n}^{c \dagger} \tilde{w}_{n}^{-} \mu_{0}^{c} + \nonumber\\
&& \displaystyle g \phi^{c}_{\nu, n} \tilde{w}_{c,n}^{-} \mu_{0} + g \mbox{\eno} \phi_{\nu,n} \tilde{w}_{c,n}^{+} \mu_{0}^{c}\\
&& \displaystyle - \frac{2m}{v} \mu_{0}^{c} \psi_{H^{+},n}^{c} \phi_{\nu,n}^{c\dagger} \nonumber
\end{eqnarray}

The external cross diagrams are of the usual form 
\begin{equation} \label{ecwino}
\left. \parbox[c]{5cm}{\includegraphics[width=4cm]{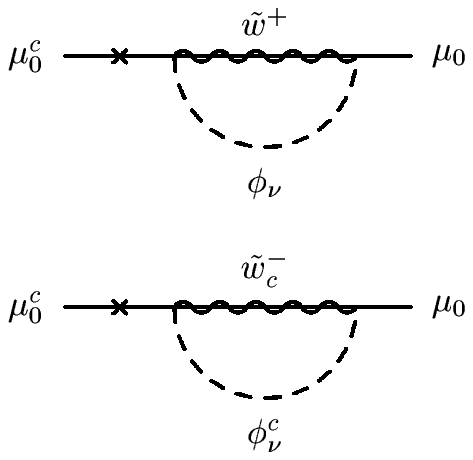}} \right\} \Rightarrow \mbox{\odd}\: x (1-x)^2 \times 2
\end{equation}
and there are three internal cross diagrams with propagators $(\tilde{w}^{+})^{\dagger}$--- $\tilde{w}_{c}^{+}$, $(\tilde{w}^{-}_{c})^{\dagger}$--- $\tilde{w}^{-}$ and $(\tilde{w}_{c}^{-})^{\dagger}$--- $\phi_{H^{+}}^{c}$:

\begin{eqnarray}
\parbox[c]{5cm}{\includegraphics[width=4cm]{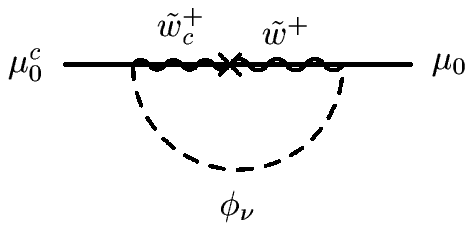}} &\Rightarrow&  \mbox{\odd}\: (1-x)^2 \label{icwino1}\\ 
\parbox[c]{5cm}{\includegraphics[width=4cm]{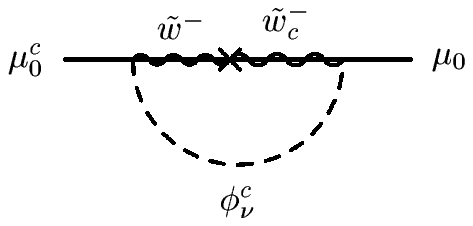}} &\Rightarrow&  - \mbox{\odd}\: (1-x)^2 \label{icwino2}\\
\parbox[c]{5cm}{\includegraphics[width=4cm]{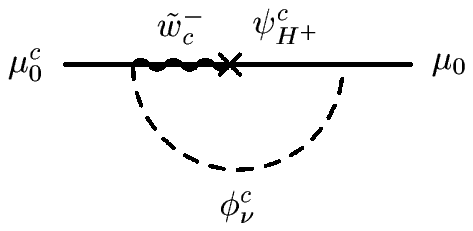}} &\Rightarrow& - \mbox{\odd}\: 4 (1-x)^2 \label{ichiggsino}
\end{eqnarray}

\subsubsection{Total charged weak contribution}

Again in this case, we correctly find that in the supersymmetric limit the sum of the KK states contributions vanishes.

We then gather all the contributions of the charged weak sector into the result:

\begin{equation} \label{finalW}
\Delta a_{\mu}^{W} = - \frac{g^2}{16 \pi^2} (m R)^2 \frac{7}{6} \frac{\pi^2}{12}
\end{equation}

\subsection{Result to leading order in $(M_W R)^2$} \label{result1}

Collecting all partial results (\ref{finalqed}), (\ref{finalZ}) and (\ref{finalW}) we have:

\begin{equation} \label{finalres}
\Delta a_{\mu}^{KK} = - \frac{g^2}{192} \frac{m^2}{M_{W}^{2}} \frac{11 -  18 s^2}{12 c^2} (M_W R)^2
\end{equation}

Numerically, for $1/R = 370\pm 70 \: GeV$, it is 

$$
\Delta a_{\mu}^{KK} = 0.07\: _{-0.02}^{+0.06} \: \cdot \: a_{\mu}^{weak} =- (1.1\: _{-0.3}^{+0.6}) \cdot 10^{-10},
$$
 which means that, for any sensible value of $R$ (see \cite{BHN-FI}), $\Delta a_{\mu}^{KK}$ is well inside the uncertainties that affect the SM prediction.

\section{$M_W R$ corrections}

In the previous section, we have neglected the ElectroWeak Symmetry Breaking in the masses of the KK modes for the gauge fields, since this was enough to get eq. (\ref{finalres}).
However, in this way we are neglecting contributions that in principle  can be quite relevant, since the dimensionless expansion parameters $M_W R$ or $M_Z R$ can be as large as 1/3.
In this chapter we estimate these contributions.
They are recovered if the full gauge fields masses are kept in the denominators of eq. (\ref{graphI})--(\ref{grV}), so the corrections to our previous results can be expressed in a simple way. Of course, these corrections only apply to the pure weak sector.

To begin with, let us focus on the neutral weak sector.
First of all, what we have to do is to study the exact spectrum: defining the dimensionless parameter $\zeta = M_Z R$, for every $n$ there is a vector boson $Z_{\mu}$ whose full mass squared is $(2n)^2 + \zeta^2$, in unit of $1/R^2$. 
The scalars fields masses are determined from eq. (\ref{Zscalarsmass}): the $n^{\mbox{th}}$ KK mode of $\Sigma_z$ has mass squared $(2n)^2 + \zeta^2$, while, for the real field $\varphi_{H^{0}}$, it has mass squared $\displaystyle (2n)^2 + \frac{3}{2} \zeta^2$.
For the neutralinos, diagonalization of eq. (\ref{neutralinosmass}) shows that there are two degenerate states with mass squared $(2n-1)^2 + \zeta^2$.
The rest of the computation proceeds as in the previous section. 
To get the modified results, it is sufficient to substitute the sum $\displaystyle \sum \frac{1}{(2n)^2}$ in eqs. (\ref{eczeta}), (\ref{iczeta}), (\ref{ecszeta}), (\ref{eczino})--(\ref{ichiggsino0}) with: 

$$
\sum \frac{1}{N^2 + \zeta^2} \quad \mbox{or} \quad \sum \frac{N^2}{(N^2+\zeta^2)^2}
$$
where $N=2n$ for bosons and $N=2n-1$ for fermions. 
The first is the one to be used in all external cross diagrams, while for internal cross ones the formulae are a bit more involved. 
To be specific, in the case of the vectors, eq. (\ref{iczeta}) becomes:

\begin{equation}
 - \frac{1}{16\pi^2} \epsilon^2 \int dx \left[ \left( g_{L} g_{R} + \frac{1}{2} (g_{L} + g_{R})^2 \right) 8 x (1-x) \frac{1}{(2n)^2 + \zeta^2}   - g_{L} g_{R}\: 16 x (1-x)^2  \frac{(2n)^2}{((2n)^2+\zeta^2)^2}  \right]
\end{equation}
With a bit of algebra, all contributions can be recast in a common form and the net result is:

\begin{equation} \label{FinalZ}
\Delta a_{\mu}^{Z} = \frac{1}{16 \pi^2} \frac{m^2}{M_W^2} (M_W R)^2 \left\{ (g_L^2 + g_R^2)\: f_1 (\zeta) + g_L g_R\: f_2 (\zeta)  \right\}
\end{equation}
where the functions $f_1$ and $f_2$ are defined in app. B.
This replaces eq. (\ref{finalZ}), with all corrections to any order in $(M_W R)^2$ included.
Numerically, for $1/R = 370 \pm 70 \: GeV$, the result is only about 3\% smaller than eq. (\ref{finalZ}). 
In the case of the charged weak sector, the calculation is very similar, so we expect the correction to be not so far from a few percent.

\section{Conclusion}

We have performed the calculation at one loop of the corrections to muon anomalous magnetic moment coming from the presence of one extra dimension in the model in ref. \cite{BHN}, to first order in $(m R)^2$ and in $(M_W R)^2$.
We have also shown that the complete calculation in $(M_W R)^2$ yields small deviations, so to our purposes the simple analytic form given in eq. (\ref{finalres}) is a good approximation.
The corrections are relatively small, for any sensible value of $R$, at the 10\% level of the pure weak contribution in the SM, and well inside the uncertainties of the hadronic contribution.

\appendix

\section*{Appendix A}

In this appendix we list the contributions to $a_{\mu}$ from the five graphs\footnote{Similar results can also be found in refs. \cite{VL}.} in figure \ref{graphsmu}:

\begin{multline} \label{graphI}
\Delta a_{\mu}^{I} = \displaystyle \frac{Q_{s}}{16 \pi^2} m^2 \int_{0}^{1} dx \left[ \frac{h_{+}^2 + h_{-}^2}{2} 2 x (1-x)^2 + \frac{m_{f}}{m} h_{+} h_{-} 2 x (1-x) \right] \frac{1}{x m_{f}^2 + (1-x) m_{s}^2 - x (1-x) m^2}
\end{multline}

\begin{multline}
\Delta a_{\mu}^{II}  =  \displaystyle - \frac{Q_{f}}{16 \pi^2} m^2 \int_{0}^{1} dx \left[ \frac{h_{+}^2 + h_{-}^2}{2} 2 x (1-x)^2 + \frac{m_{f}}{m} h_{+} h_{-} 2 (1-x)^2 \right]\\ \frac{1}{x m_{s}^2 + (1-x) m_{f}^2 - x (1-x) m^2}
\end{multline}

\begin{multline}
\Delta a_{\mu}^{III}  =  \displaystyle \frac{Q_{v}}{16 \pi^2} m^2 \int_{0}^{1} dx \left[\frac{g_{+}^2 + g_{-}^2}{2} 2 (1-x)^2 (2x-3) + g_{+} g_{-} \frac{m_{f}}{m} 6 (1-x)^2 \right]\\ \frac{1}{x m_{f}^{2} + (1-x) m^{2}_{v} - x(1-x) m^2}
\end{multline}

\begin{multline} \label{graphIV}
\Delta a_{\mu}^{IV} = - \displaystyle \frac{Q_{f}}{16 \pi^2} m^2 \int_{0}^{1} dx \left[ \frac{g_{+}^2 + g_{-}^2}{2} \left( 4 x (1-x)^2 - 8 x (1-x) \right) + \frac{m_{f}}{m} g_{+} g_{-} 8 x (1-x) \right]\\ \frac{1}{x m_{v}^2 + (1-x) m_{f}^2 - x (1-x) m^2}
\end{multline}

\begin{equation}
\label{grV}
\Delta a_{\mu}^{V} = \displaystyle \frac{Q_{v}}{16 \pi^2} m \tau \int_{0}^{1} dx (g_{+} h_{+} + g_{-} h_{-} ) (1-x)^2 \frac{1}{x m_{f}^2 + (1-x) m_{v}^2 - x (1-x) m^2}
\end{equation}
where $Q_{s,f,v}$ are the electric charges of the particles emitting the zero-mode photon and $m_{s,f,v}$ are the masses in the loop.
Note that in eq. (\ref{grV}) we assume $m_s = m_v$.
The couplings are defined by:
\begin{eqnarray*}
& v_{\rho} \overline{f} \gamma^{\rho} \left[ g_{+} \mathcal{P}^{+} + g_{-} \mathcal{P}^{-} \right]  \mu &\\
& s \overline{f} \left[ h_{+} \mathcal{P}^{+} + h_{-} \mathcal{P}^{-} \right]  \mu &\\
& e \tau  s^{+} W^{-}_{\rho} A^{\rho} &\\
& e \left[ g^{\nu \rho} ( k - p_{+})^{\sigma} + g^{\rho \sigma} (p_{+} - p_{-})^{\nu} + g^{\sigma \nu} (p_{-} - k)^{\rho} \right] W^{+}_{\rho} W^{-}_{\sigma} A_{\nu} &
\end{eqnarray*}
where $P^{\pm}$ are projectors on the two Weyl components of the Dirac fields.

\section*{Appendix B}

In this appendix we list the functions to be inserted in eq.  (\ref{FinalZ}):

$$
\begin{array}{rcl}
f_1 &=& \displaystyle \frac{1}{\zeta^8} \int_{0}^{\zeta} dy \: 2 (\zeta^4-y^4) \left(\pi y^2 \mbox{tanh} \left(\frac{\pi}{2} y \right) - \pi y^2 \mbox{coth}\left(\frac{\pi}{2} y\right) + 2 y\right) =\\
 &=& \displaystyle \frac{4}{3 \zeta^2} + \frac{1}{3 \pi^6 \zeta^8} \bigg[ 48 \pi^5 \zeta^5 \mbox{\li{2}} + 336 \pi^4 \zeta^4 \mbox{\li{3}}   \\
& & \displaystyle+ 1440 \pi^3 \zeta^3 \mbox{\li{4}} +  4320 \pi^2 \zeta^2 \mbox{\li{5}} + \\
 & & \displaystyle + 8640 \pi \zeta \mbox{\li{6}} + 8640 \mbox{\li{7}} +\\
 & & \displaystyle  - 42 \pi^4 \zeta^4 \: \zeta(3) + 17145 \; \zeta(7) \bigg] 
\end{array}
$$

$$
\begin{array}{rcl}
f_2 &=& \displaystyle \frac{1}{\zeta^8} \int_{0}^{\zeta} dy \: 8 (y^2-\zeta^2) (3y^2-2\zeta^2) \left(\pi y^2 \mbox{tanh} \left(\frac{\pi}{2} y\right) - \pi y^2 \mbox{coth}\left(\frac{\pi}{2} y\right) +2 y\right) =\\
 &=& \displaystyle \frac{4}{\zeta^2} - \frac{4}{\pi^6 \zeta^8} \bigg[ 8 \pi^5 \zeta^5 \mbox{\li{2}} + 136 \pi^4 \zeta^4 \mbox{\li{3}} + \\
 & & \displaystyle +960 \pi^3 \zeta^3 \mbox{\li{4}}  + 3840 \pi^2 \zeta^2 \mbox{\li{5}} + \\
& & \displaystyle + 8640 \pi \zeta \mbox{\li{6}} + 8640 \mbox{\li{7}} +\\
& & \displaystyle+ 28 \pi^4 \zeta^4 \: \zeta(3) - 930 \pi^2 \zeta^2 \: \zeta(5) + 17145\:  \zeta(7) \bigg]
\end{array}
$$
where $\zeta (n)$ is the Riemann zeta function, not to be confused here with our variable $\zeta = M_Z R$.
In the integrals, the hyperbolic tangent is produced by the sum over the fermionic KK states, while the cotangent and the ``$2y$'' pieces by the bosonic states.

\section*{Acknowledgements}

We would like to thank  Riccardo Barbieri and Riccardo Rattazzi for useful discussions and suggestions.
We are also grateful to Gilberto Colangelo for informations about the present status of SM $(g-2)_{\mu}$.
This work was supported by the EC under the RTN contract HPRN-CT-2000-00148. 

\pagebreak

%%%%%%%%%%%%%%%%%%%%%%%%%%%%%%%%%%%%%%%%%%%%%%%%%%%%%%%%%%%%%%%%%%%%%%%%
%%%%%%%  BIBLIO    %%%%%%%%%%%%%%%%%%%%%%%%%%%%%%%%%%%%%%%%%%%%%%%%%%%%%
%%%%%%%%%%%%%%%%%%%%%%%%%%%%%%%%%%%%%%%%%%%%%%%%%%%%%%%%%%%%%%%%%%%%%%%%

\end{document}